\documentclass[graybox]{svmult}
\usepackage{mathptmx}       % selects Times Roman as basic font
\usepackage{helvet}         % selects Helvetica as sans-serif font
\usepackage{courier}        % selects Courier as typewriter font
\usepackage{type1cm}        % activate if the above 3 fonts are
\usepackage{amsfonts}
\usepackage[english]{babel}
\usepackage{amssymb}
\usepackage{amscd}
\usepackage{mathrsfs}
\usepackage{amsmath}
\usepackage{amscd}
\newcommand{\be}{\begin{equation}}
\newcommand{\ee}{\end{equation}}
\newcommand{\beqa}{\begin{eqnarray}}
\newcommand{\eeqa}{\end{eqnarray}}

\usepackage{graphicx}        % standard LaTeX graphics tool
\usepackage{multicol}        % used for the two-column index
\usepackage[bottom]{footmisc}% places footnotes at page bottom

\begin{document}

\title*{Construction of the noncommutative rank I Bergman domain}
\author{Zhituo Wang}
\institute{ Laboratoire de Physique Th\'eorique, CNRS UMR 8627,\\
Universit\'e Paris XI,  F-91405 Orsay Cedex, France, \email{zhituo.wang@th.u-psud.fr}}
\maketitle

\abstract{
In this paper we present a harmonic oscillator realization of the
most degenerate discrete series representations of the $SU(2,1)$
group and the deformation quantization of the coset space
$D=SU(2,1)/U(2)$ with the method of coherent state quantization. This short article is based on a talk given at the 9-th International Workshop, Varna 
"Lie Theory and Its Applications in Physics" (LT-9).}

\section{Introduction}
It is believed that ordinary differential geometry should be replaced by noncommutative
geometry \cite{connes} when we are approaching the Planck scale and quantum field theories defined on
noncommutative space time (NCQFT) \cite{Douglas, Szabo, Bourbaphy} are considered as the right way to explore the effects of quantum gravity.

The simpliest noncommutative space is the Moyal space, which is a symplectic manifold
generated by the noncommutative coordinates $x_\mu$, such that  $[x_\mu, x_\mu]=i\theta_{\mu\nu}$, where $\theta_{\mu\nu}$ is a constant. The first well defined quantum field theory on 4 dimensional Moyal space is the Grosse-Wulkenhaar model \cite{GW1}. It is not only perturbative renormalisable to all orders but also asymptotically safe, namely the beta function for the coupling constant is zero at the fixed point of this model. Hence this model is a candidate to be constructed nonperturbatively, namely it's possible to obtain the exact Green's function which is unique and analytic in the coupling constant, by resumming the perturbation series \cite{rivass}. Recently the two dimensional Grosse-Wulkenhaar model has been constructed in \cite{Wanggw}.

Since the noncommutative quantum fields theories are better behaved than their commutative counterparts,
it is very natural to construct other noncommutative manifolds and physics models over them.

In this paper we construct the noncommutative coset space $D=SU(2,1)/S(U(2)\times U(1))$, with the method 
of coherent state quantization. For doing this we also introduce a harmonic oscillator realization of the 
most degenerate discrete series representation
of the group $SU(2,1)$ which is a generalization for the $SU(1,1)$ case introduced by H. Grosse and P. Presnajder \cite{GP}. The interested reader could look at \cite{GP,Perelomov} for more details about the coherent state quantization and \cite{rudin,Kirillov,knapp} for more details about the representation theory of noncompact Lie group. In  \cite{bergman1} and \cite{Wang1} we have studied the harmonic oscillator realization of the maximal degenerate discrete series representations for an arbitrary $SU(m,n)$ group.

The construction of the noncommutative coset space $SU(2,1)/U(2)$ has been also studied
by \cite{Upmeier}, \cite{Jakim} with the method of Berezin-Toeplitz quantization and by
by \cite{BGR} with the method of "WKB quantization". The interested
reader could go to the references for details.

\section{The $SU(m,1)$ group and its Lie algebra}
The group $G= SU(m,1)$ is defined as a subgroup of the matrix group
$SL(m+n,C)$: 
\begin{equation}\label{def}
   g=\left(\begin{array}{cc}a_{m\times m}&b_{m\times 1}\\ c_{1\times m}& d\end{array}\right)\in G
\end{equation}
satisfies the constraint
\begin{equation}\label{unit}
g^\dagger \Gamma g=g \Gamma g^\dagger=\Gamma ,\ \ \
%\end{equation} where\begin{equation}\label{gamma}
\Gamma=\left(\begin{array}{cc}{I_{m\times m}}&0\\ 0& -1\end{array}\right).
\end{equation}
Here $I$'s represents unit matrices and $0$'s the blocks of zeros.

The maximal compact subgroup is defined by matrices
\begin{equation}
K=S(U(m)\times U(1))=\{\left(\begin{array}{cc}K_1&0\\0&K_2\end{array}\right),\ \det(K_1 K_2)=1\}.
\end{equation}

% The Lie algebra $\textbf{g}=Lie(G)\,=\,su(m,1)$ is defined by
% \be
% M=\big\{\left(\begin{array}{cc}A&B\\ B^\dagger&
% D\end{array}\right) \big\}\in \textbf{g}, \ M^\dagger\Gamma=-\Gamma M,
% \ee
% where
% $A^\dagger\,=\,-A, D^\dagger\,=\,-D, \mbox{tr}(A+D)\,=\,0$.

The Bergman domain is defined as the coset space $D=G/K$: 
\begin{equation}
D=\{Z\,|\, 1-|Z|^2 > 0\}\ =\ \{z\,|\, 1-|z_1|^2-|z_2|-\cdots
-|z_m|^2
> 0\},
\end{equation}
where $Z=(z_1,\cdots, z_m)$ are the coordinates of the coset space $D$.
It is a pseudo-convex domain over which we could define a holomorphic
Hilbert spaces \cite{Upmeier} with the reproducing Bergman kernel:
\begin{equation}
K(W^\dagger, Z)=(1-W^\dagger Z)^{-N},
\end{equation}
where $Z$ and $W$ are complex m-columns, and $N=m+1,m+2,\cdots$ is a
natural number characterizing the representation.

$D$ is also a K{\" a}hler manifold with
the K{\"a}hler metric defined by the derivations of the Bergman kernel:
\begin{equation}
 g_{i\bar j}\ =\frac{1}{N} \partial_{\bar{z}^i}\partial_{z^j}\,\log K(Z^\dagger,Z).
\end{equation}
More explicitly we have:
\begin{equation}
 g_{i \bar j}=[\frac{\delta_{ij}}{1-|Z|^2}+\frac{z_i\bar{z}_j
 }{(1-|Z|^2)^2}], \  \  g^{i \bar j}=(1-|Z|^2)(\delta_{ij}- \bar z_i z_j).
\end{equation}

We could easily calculate that the Ricci tensor: $R_{i\bar j}=-(m+1)g_{i\bar{j}}$ and the curvature $R=-(m+1)$
and  verify that the metric $g_{i\bar j}$ is a
solution to the Einstein's equation in the vacuum:
\begin{equation}\label{Eevb}
R_{i\bar j}-\frac{1}{2}g_{i\bar j} R+\Lambda g_ {i\bar j}=0
\end{equation}
with the cosmological constant $\Lambda=\frac{m+1}{2}$.

%\subsection{The Lie algebra}
The Lie algebra $\textbf{g}=Lie(G)\,=\,su(m,1)$ is defined by
$M=\big\{\left(\begin{array}{cc}A&B\\ B^\dagger&
D\end{array}\right) \big\}\in \textbf{g}$, $M^\dagger\Gamma=-\Gamma M$.
where
$A^\dagger\,=\,-A, D^\dagger\,=\,-D, \mbox{tr}(A+D)\,=\,0$.

Consider the Cartan decomposition of the Lie algebra $\textbf{g}=\textbf{l}+\textbf{p}$ and
let $\textbf{a} \in \textbf{p}$ be a maximal Abelian subalgebra. We
could choose for $\textbf{a}$ the set of all matrices of the form
\begin{equation}
H_t=\begin{pmatrix}
O_{(m-1)\times(m-1)} & O_{(m-1)\times 1} & O_{(m-1)\times 1} \\
O_{1\times(m-1)} & 0 & t \\
O_{1\times(m-1)} & t & 0 \\
\end{pmatrix}
\end{equation}
where $t$ is a real number.

Define the linear functional over $H_t$ by $\alpha(H_t)=t$, the
roots of $(\textbf{g},\textbf{a})$ are given by
\begin{equation}
\pm \alpha,\  \pm 2\alpha,
\end{equation}
with multiplicities $m_{\alpha}=2$ and $m_{2\alpha}=1$.

Define
\begin{equation}
\delta:= \{\textbf{a}_t|\  \textbf{a}_t=\exp H_t,\ H_t\in
\textbf{a}\}.
\end{equation}
so we have
\begin{equation}
\textbf{a}_t=\begin{pmatrix}
I & O & 0\\
O & \cosh t & \sinh t \\
0 & \sinh t& \cosh t \\
\end{pmatrix},
\end{equation}
where the symbol $I$ stands for the identity matrix and $O$ is the matrix with entries
$0$.
\section{The holomorphic discrete series of representations of the $SU(m,1)$ group}
The unitary irreducible representations for $G=SU(m,1)$ are the principal series, the discrete series and the
supplementary series. We consider only the discrete series of
representations, which are realized in the Hilbert
space $\mathcal {L}^2_N(D)$ of holomorphic functions with the 
the inner product defined by:
\be
(f,g)_N=\int d\mu_N(Z,\bar Z)\ \bar f(\bar Z) g(Z),
\ee
where $d\mu_N(Z,\bar Z)= c_N [\det(E-Z^\dagger Z)]^{N-(m+1)}|dZ|$ is the normalised measure and
$c_N=\pi^{-2}(N-2)(N-1)$.

The discrete series of representations $T_N$ is defined by
\begin{equation}\label{rep}
T_{N}f(Z)=[\det(CZ+d)]^{-N}f(Z'),  \ N=m+1, m+2,\cdots
\end{equation}
where
\begin{equation}
Z'=(AZ+B)(CZ+d)^{-1}
\end{equation}

In the following we consider only the $m=2$ case and construct the harmonic oscillator realization of the most degenerate discrete series representation \eqref{rep}.
We introduce a $3\times 1$ matrix $\hat{Z} = (\hat{z}_{a})$,
$a\,=\,1, 2, 3$, of bosonic oscillators
acting in Fock space and satisfying commutation relations
\begin{equation}\label{commu1}
 [\hat{z}_{a},\hat{z}^\dagger_{b}]=\Gamma_{ab},\  a, b=1, 2 , 3
\end{equation}
\begin{equation}
[\hat{z}_a,\hat z_b]=[\hat{z}^\dagger_a,\hat z^\dagger_b]=0,
\end{equation}
where $\Gamma$ is a $3\times 3$ matrix defined in (\ref{unit}). It
can be easily seen that for all  $g\,\in\,SU(2,1)$ these commutation
relations are invariant under transformations:
\be\label{2.1au1} \hat{Z}\ \mapsto\ g\,\hat{Z},\ \ \ \hat{Z}^\dagger\
\mapsto\ \hat{Z}^\dagger\,g^\dagger .\ee
We could define the creation and anhilation operators $\hat a_\alpha$ and $\hat b$ as:
\be\label{commu2}  \hat{Z}\ =\ \left(\begin{array}{c}\hat{a}\\
\hat{b}^\dagger\end{array}\right):\ \
[\hat{a}_{\alpha},\hat{a}^\dagger_{\beta}]\ =\
\delta_{\alpha\beta}, \alpha,\beta=\,1,2.  \ \
[\hat{b},\hat{b}^{\dagger}]\ =\ 1,
 \ee
and all other commutation relations among oscillator operators
vanish. 

The Fock space  ${\cal F}$ in question is generated from a
normalized vacuum state $|0\rangle$, satisfying
$\hat{a}_{\alpha}\,|0\rangle\,=\,\hat{b}\,|0\rangle\,=\,0$,
by repeated actions of creation operators:
\be\label{fock} |m_{\alpha},\,n\rangle\ =\
\prod_{\alpha}\,\frac{(\hat{a}^\dagger_{\alpha})^{m_{\alpha}}\,(\hat{b}^\dagger)^
n}{\sqrt{m_{\alpha}!\,n!}}\,|0\rangle\,.\ee
\vskip0.5cm
We shall use the terminology that the state
$|m_{\alpha},\,n\rangle$ contains $m=\sum
m_{\alpha}$ particles  $a$ and $n$
particles  $b$.

Consider a basis of $su(2,1)$ Lie algebra
$X\,=\,X^{A}_{ab}$ $A=1, \cdots 8$, $a,b=1,2,3$, we assign the operator
\be\label{2.3au1} \hat{X}\,=\,-\hat{Z}^\dagger\Gamma
X\hat{Z}\ =\
-\hat{z}^\dagger_{a}\,\Gamma^{A}_{ab}\,X^A_{ab}\,\hat{z}_{b}\, .\ee

Their anti-hermicity follows directly:
$$ \hat{X}^\dagger\ =\ -\mbox{tr}\,(\hat{Z}^\dagger
X^\dagger\Gamma \hat{Z})\ =\ +\,\mbox{tr}\,(\hat{Z}^\dagger\Gamma
X\hat{Z})\ =\ -\,\hat{X}. $$

Using commutation relations for annihilation and creation operators we have:
\be\label{2.4u1} [\hat{X},\hat{Y}]\ =\
[\hat{Z}^\dagger\Gamma
X\hat{Z},\hat{Z}^\dagger\Gamma Y\hat{Z}]\ =\
-\hat{ Z}^\dagger\,\Gamma\,[X,\ Y]\,\hat{Z}\,.\ee
%%%%%%%%%%%%%%%%%%%%%%%%%%%%%%%%%
So that the operators $\hat{X}_{a}$ satisfy in Fock space the $su(2,1)$ commutation
relations.  The assignment
 \be\label{2.4au1}
g\,=\,e^{\xi^{A}X_{A}}\,\in\,SU(2,1) \Rightarrow\
\hat{T}(g)\,=\,e^{\xi^{A}\hat{X}_{A}} \ee
then defines a unitary $SU(2,1)$ representation in Fock space.

% We start to construct the representation space ${\cal F}_N$ from a distinguished normalized state containing lowest number of particles:
%  %c^{-1}_N\,=\,N!\,\sqrt{N+1},
%  

\section{The Coherent States Quantization of $D=SU(2,1)/U(2)$ and the Star Product}
We briefly describe the construction of coherent states on coset space of a Lie group following  \cite{Perelomov}.
% We choose a set of boosts of the form:
% %
% \begin{equation}\label{2.11} g_z =  k\,\delta\,k^\dagger\ =\begin{pmatrix}
% C'&S'\\S'^\dagger&C''\\ \end{pmatrix}\in G/K,
% \end{equation}
% where $C'=k'\begin{pmatrix}1&0\\0&\cosh t\\  \end{pmatrix}k'^\dagger
% $, \ $S'=\begin{pmatrix} k'_{12}\sinh t\\k''_{22}\sinh t\\  \end{pmatrix}k''^\star
% $,\  $C''=\cosh t$. $k=diag(k',k'')\in K$.
Let $T_g$ be an unitary irreducible representation of an arbitrary Lie group $G$ in a Hilbert space $\mathcal{H}$,
$|z_0\rangle\in\mathcal{H}$ is a normalized state in the Garding space of $T_g$.
%\item The UIR $T_g$ on the the Hilbert space of states $\psi\mathcal{H}$  :
Let $K$ be the stability group of the $|z_0\rangle$, for  which 
$T_k|z_0\rangle=e^{i\alpha(k)}|z_0\rangle,\ \ {\rm for}\  k\in K$. Then for each $z=g_z z_0\in D=G/K$ we could assign a coherent states :
$ |z\rangle=\psi_z= T(g_z)|z_0\rangle$.
Define the functions $\omega_0(g)=<z_0|T(g)|z_0>$ and $\omega (g,z)=<z|T_g|z>=\omega_0 (g_z^{-1}g g_z) $. As $|z_0\rangle$ is in the Garding space, $\omega(g)$ is a smooth function in $g$.

For $G=SU(2,1)$, the state $ |z_0\rangle$ is defined in the Fock space as:
\be\label{2.7u1} |z_0\rangle = \frac{(\hat{b}^\dagger)^{N}
 }{\sqrt{(N)!}}|0> =
 \frac{1}{\sqrt{N}}\,|0,0;N\rangle\,.\ee
Here $N$ is a natural number that specifies the representation: $ \hat{N}\,|z_0\rangle\,=\,N\,|z_0\rangle$. All other states in the representation space are obtained by the action of rising operators given in (\ref{fock}). The stability group for $|z_0\rangle$ is $K\,=\,S(U(2)\times U(1))$.

Using the $K\delta K$ decomposition of $g = k^{\dagger}\,\delta\,q$ \cite{knapp}, for which $k,q\in K$, we obtain:
\be\label{2.10} \omega_0(g) = \langle z_0|\,\hat{T}(g)\,|z_0\rangle
= \frac{1}{\cosh t}[(1+\ln\cosh t)e^{i(\alpha(q)-\alpha(k))}]^N \ee.

% To any boost $g_z\in G/K$ we assign a coherent state by:
% %
% \be\label{2.12} |z\rangle\ =\  \hat{T}(g_z)\,|z_0\rangle\ =\
% \hat{T}(k\,\delta\,k^\dagger)\,|z_0\rangle,\ \   z = z(k,\delta).
% \ee
%
Consider an operator acting on $\mathcal{H}$:
\begin{eqnarray}\label{star}
\hat F&=&\int dg \tilde F(g)T(g)=\int dg \tilde F(g)\omega( g, z)
\end{eqnarray}
where $\tilde{F}(g)$ ia a distribution on a group $G$ with compact
support. We also define for each $\hat F$ a biholomorphic function:
\begin{equation}
F(z,\bar z)=\langle \Psi_z|\hat F|\Psi_z\rangle.
\end{equation}
The star product of two such bi-holomorphic functions $F$ and $G$ is then defined by \cite{GP}:
\begin{eqnarray}\label{star1}
&&(F\star G)(z,\bar z)=\langle \Psi_z|\hat F\hat G|\Psi_z\rangle=(F\star G)(z,\bar z)\nonumber\\
&=&\int dg_1 dg_2 \tilde F(g_1) \tilde
F(g_2)\omega( g_1 g_2, z).
\end{eqnarray}
% This equation combined with (\ref{2.10}) offers an explicit form of
% $\omega(g,x)$ and is well suited for calculations.

% The star-product of two functions is defined by \cite{GP}:
% %
% $$ (F\star G)(z)\ =\ \langle z|\,\hat{F}\hat{G}\,|z\rangle\ =\ \int_{G\times G} dg_1 dg_2\,\tilde{F}(g_1)\,\tilde{G}(g_2)\, \omega(g_1g_2,z)$$.
% 
Obviously the star product defined above is noncommutative, associative and is invariant
under the action of the group $G$. The noncommutative algebra of functions $\{F(z)\}$
induces a noncommutative structure on the coset space $D$. That's how we construct the noncommutative version
of the Bergman domain, which is noted as $\hat D$.

Now we shall study the explicit form of the star product for $G=SU(2,1)$. Using the explicit form of the group element $g =
e^{\xi^{A} X_{A}}$ and integration by parts we have:
\be\label{2.20} F_{A_1\dots A_n}(z)\ =\ (-1)^n (\partial_{\xi_{A_1}}\,\dots\, \partial_{\xi_{A_n}}\omega)( e^{\xi^{A}\hat X_{A}}z)|_{\xi=0}=
(-1)^n\,\langle z|\,\hat{X}_{A_1}\,\dots\,
\hat{X}_{A_n}\\|z\rangle.\ee

Here ${\hat X}_{A}$, $(A=1\cdots, 8)$ are the left-invariant vector field on group $G$
corresponding to the Lie algebra basis $X_{A}$ whose explicit form is given in \cite{Wang1}.
% Here we used the fact that the operators ${\cal X}_{A}$ are
% anti-hermitian differential operators with respect to the group
% measure $dg$. 

From the definition of the star product \eqref{star1} it follows that:
$$ (F_{A_1\dots A_n}\star F_{B_1\dots B_m})(z) $$
\be\label{2.19}   =\ (-1)^{n+m}({\hat X}_{A_n}\,\dots\, {\hat
X}_{A_1}\,{\hat X}_{B_m}\,\dots\, {\hat
X}_{B_1}\omega)(g,z)|_{g=e}.\ee
\vskip0.5cm
% 

%
%For $G=SU(2,1)$ case we have $K=$ 

We define the function $\xi_A$ as the expectation value of the operator $\hat{X}_A$ between the coherent states as:
\be\label{nccoord} \xi_{A}(z) = \frac{1}{N}\,\langle
z|\hat{X}_{A}|z\rangle = \frac{1}{N}\,\langle z_0|\hat{T}^\dagger
(g_z)\,\hat{X}_{A}\hat{T}(g_z)|z_0\rangle. \ee

The star product between these coordinates functions reads:
% \be\label{starcoo}
% \xi_A \star \xi_B=\frac{1}{N^2}\,\langle
% z|\hat{X}_{A}\hat{X}_{B}|z\rangle,
% \ee
% 
% 
% And we have:
\be\label{star2} (\xi_{A}\star\xi_{B})(z)\ =\frac{1}{N^2}\,\langle
z|\hat{X}_{A}\hat{X}_{B}|z\rangle=\ (1+ A_N)\,
\xi_{A}(z)\,\xi_{B}(z) \,+\,\frac{1}{2N}\,f^{C}_{A,B}
\,\xi_{C}(z)\,+\,B_N\, \delta_{A,B}, \ee
where $A_N$ and $B_N$ depend on the Bernoulli numbers coming from the Baker-Campbell-Hausdorff
formula and are of order $1/N$. We see that the parameter of the non-commutativity is $\lambda_N =
1/N$. For $N\,\to\,\infty$ we recover the commutative product.

According to the Harish-Chandra imbedding theorem, we could always
imbed the commutative maximal Hermitian symmetric space into the noncompact part of
the Cartan subalgebra. So the coordinates of the noncommutative Bergman domain $\hat D$ can be identified as the coordinate functions corresponding to the noncompact
Cartan subalgebra.

\section{Conclusions and prospectives}
In this paper we have constructed the noncommutative Bergman domain $\hat D$ whose commutative
counterpart is the coset space $D=G/K$, where $G=SU(2,1)$ and $K=S(U(2)\times U(1))$. This result
could be generalized to an arbitrary type one rank one Cartan domain $D=G=SU(m,1)/S(U(m)\times U(1))$
straightforwardly.

In \cite{Wang1} we have build a model of quantum theory of real scalar fields on the noncommutative manifold $\hat D$ and find that
the one loop quantum correction to the 2 point function is finite. This is a hint of the finiteness of quantum field theory on $\hat D$ and this
deserves further studies.

{\bf Ackonledgements}\  The author is very grteful to
Harald Grosse and Peter Presnajder for useful discussions.

\end{document}